\documentclass[aps,prd,twocolumn,letterpaper,superscriptaddress, showpacs, amsmath, amssymb]{revtex4}
\usepackage{graphicx}
\usepackage{amsmath,amssymb}
\usepackage{url}
\usepackage{color}

\def\beq{\begin{equation}}
\def\enq{\end{equation}}
\def\bea{\begin{eqnarray}}
\def\ena{\end{eqnarray}}

\def\Ek525{E_{k,52.5}}

\def\tev{\mathrm{TeV}}
\def\pev{\mathrm{PeV}}
\def\llgrb{LL GRBs}

\begin{document}


\title{Choked Jets and Low-Luminosity Gamma-Ray Bursts as Hidden Neutrino Sources}

\author{Nicholas Senno}
\author{Kohta Murase}
\author{Peter M\'esz\'aros}
\affiliation{Department of Physics; Department of Astronomy \& Astrophysics; Center for Particle and
Gravitational Astrophysics, The Pennsylvania State University, University Park, PA 16802, USA}

\date{\today}

\begin{abstract}
We consider gamma-ray burst (GRB) jets that are choked by extended material as sources of high-energy cosmic neutrinos.  
We take into account the jet propagation physics both inside the progenitor star and the surrounding dense medium. 
Radiation constraints, which are relevant for high-energy neutrino production are considered as well.  Efficient shock acceleration 
of cosmic rays is possible for sufficiently low-power jets and/or jets buried in a dense, extended wind or outer envelope. Such conditions 
also favor GRB jets to become stalled, and the necessary conditions for stalling are explicitly derived.
Such choked jets may explain transrelativistic supernovae (SNe) and low-luminosity (LL) GRBs, giving a unified picture of GRBs and GRB-SNe.  
Focusing on this unified scenario for GRBs, we calculate the resulting neutrino spectra from choked jets including the relevant microphysical processes 
such as multipion production in $pp$ and $p\gamma$ interactions, as well as the energy losses of mesons and muons. 
We obtain diffuse neutrino spectra using the latest results for the luminosity function of LL GRBs.  Although uncertainties are large, 
we confirm that LL GRBs can potentially give a significant contribution to the diffuse neutrino flux. Our results are consistent with the present IceCube data and do not violate the stacking limits on classical high-luminosity GRBs.
We find that high-energy neutrino production in choked jets is dominated by $p\gamma$ interactions. 
These sources are dark in GeV-TeV gamma-rays, and do not contribute significantly to the {\it Fermi} diffuse gamma-ray background.  
Assuming stalled jets can launch a quasi-spherical shock in the dense medium, ``precursor'' TeV neutrinos emerging prior to the shock breakout 
gamma-ray emission can be used as smoking gun evidence for a choked jet model for LL GRBs.  
Our results strengthen the relevance of wide field-of-view sky monitors with better sensitivities in the $1-100$~keV range.
\end{abstract}

\pacs{95.85.Ry, 97.60.Bw, 98.70.Rz\vspace{-0.3cm}}

\maketitle


\section{Introduction}
\label{sec:intro}
A diffuse flux of very high-energy (VHE) neutrinos with energies $10 \,\tev \lesssim E_\nu \lesssim 2 \,\pev$ has been reported from the Antarctic neutrino detector IceCube \citep{Aartsen:2013bka,Aartsen:2013jdh,Aartsen:2014gkd,Aartsen:2014muf,Aartsen:2015ita,Aartsen:2015rwa,Botner:2015ipa,Ishihara:2015tev}. The sources of these neutrinos are currently unknown, but they appear to be extragalactic in origin \citep{Waxman:2013zda,Meszaros:2014tta,Murase:2014tsa}. 
Gamma-ray bursts (GRBs), energetic supernovae (SNe), active galactic nuclei (AGNs), starburst and star-forming galaxies, galaxy groups and clusters, as well as some Galactic sources have been proposed as potential candidates (see reviews~\citep{Waxman:2015ues,Meszaros:2015krr,Murase:2015ndr}). 
In particular, GRBs, which are believed to be caused by ultra-relativistic jets launched by the collapse of a massive star (i.e. collapsars or long GRBs) or the merger of two compact binary objects (e.g. \cite{2006RPPh...69.2259M} for review) have been investigated as potential sources of ultrahigh-energy cosmic rays (CRs) and secondary VHE neutrinos for more than a decade~\citep{Waxman:1997ti}. Stacking analyses \cite{Abbasi:2012zw,Aartsen:2014aqy} lead to the conclusion that $\lesssim1$\% of the measured diffuse flux can be explained as prompt emission from observed high-luminosity (HL) GRBs~\cite{Murase:2013ffa,Laha:2013lka,Liu:2012pf}.  
However, low-power GRBs \footnote{``Low-power GRBs'' here are introduced based on the observed luminosity. Even if the observed luminosity of LL GRBs is $L_\gamma\sim{10}^{47}~{\rm erg}~{\rm s}^{-1}$, which is much lower, the intrinsic isotropic-equivalent luminosity of choked jets may be as high as that of classical GRBs.}, including low-luminosity GRBs (\llgrb) and ultralong GRBs (UL GRBs), may be largely missed by current GRB satellites such as {\it Swift} and {\it Fermi}.  LL GRBs may be more common than classical GRBs \cite{Liang:2006ci,howell+13pop,sun+15rates} and UL GRBs may also be as common as high-luminosity GRBs \cite{levan+14ulgrb}. 
In addition, choked jets that do not escape their progenitor star in the collapsar scenario -- so called failed GRBs~\cite{Meszaros:2001ms}  -- have been suggested as possible sources for the observed diffuse neutrino flux~\cite{Murase:2013ffa,Murase:2015xka,Bhattacharya:2014sta}.  
The present stacking limits on prompt neutrino emission from classical high-luminosity GRBs are not applicable to such low-power or dark GRBs, and these may give a significant contribution to the diffuse neutrino flux~\cite{Murase:2006mm,Gupta:2006jm}. 
For the first time, we study the expected neutrino contribution from jets that successfully escape from their progenitor star, but are subsequently smothered by a dense, optically thick external material resulting in a LL GRB or a failed GRB. 
This scenario, described by Nakar~\cite{nakar15-060218}, allows for a unified picture for HL, LL, and dark GRBs which have similar intrinsic progenitor and jet properties, but different circumstellar environments (e.g. the presence or absence of a dense wind or outer envelope). It is conjectured that LL GRBs may occur if the relativistic jet becomes smothered by the extended wind/outer envelope, and acts as a piston driving a quasi-spherical shock into the circumstellar material. The GRB event occurs when this shock breaks out in the optically-thin region.

Another possible subclass of interest are UL GRBs, which have a much longer duration compared to classical GRBs (but see also Ref.~\cite{Zhang:2013coa}).  Their long duration may suggest a long-lasting fall-back accretion from an extended progenitor onto a black hole.  Blue supergiants (BSGs) are possible UL GRB progenitors, and are believed to be common at very high redshifts~\cite{meszaros+rees10pop3,nakauchi+13bsg}.
Alternatively, such long durations may be explained by a fast-rotating pulsar, which could account for the connection between UL GRBs, super-luminous SNe and hypernovae~\citep[e.g.,][]{Thompson:2004wi,greiner+15ulgrb,kashiyama+15mmsne}. Although we do not consider potential sources of UL GRBs in this work, these low-power GRBs can also
contribute to neutrino emission~\cite{Murase:2013ffa}.

Predictions for high-energy neutrino emission from GRB jets of both high and low luminosity are still uncertain despite recent improvements in theoretical calculations~
\citep[e.g.,][]{Murase:2005hy,Murase:2008sp,He:2012tq,Hummer:2011ms,Gao:2012ay,Asano:2014nba,Bustamante:2014oka} (although guaranteed emission is expected in the GeV-TeV range for neutron-loaded outflows~\citep[e.g.,][]{Bahcall:2000sa,Meszaros:2000fs,Murase:2013hh,Kashiyama:2013ata}). 
Irrespective of their viability as VHE neutrino factories, the mechanisms for producing, and the physical processes associated with low-power GRBs are still largely unknown and remain intriguing open questions. Nearby long GRBs have been associated with broad-line Type Ic supernovae (SNe) (e.g., GRB 980425, 060218, and 100316D), which are known to be caused by the collapse of massive stars that eject of their outer envelopes.
LL GRBs have been of special interest since they show intermediate properties between GRBs and SNe, and have been associated with transrelativistic SNe \cite{soderberg+06-060218}. Both types of transients may be driven by jets \cite{nakar15-060218,irwin+chevalier15jets} and the study of LL GRBs may offer clues to the GRB-SN connection \cite{margutti+13grb-sn,Margutti:2014gha}. 

In this work, based on the above motivation we consider the VHE neutrino emission from jets choked by dense external material, as well as any subsequent shocks resulting from the jet acting as a relativistic piston. In particular, we focus on scenarios which may produce LL GRBs. Under the current constraints imposed by the IceCube analyses mentioned above, such LL GRBs are attractive as the originators of the diffuse VHE neutrino flux (i) because of their high local rate relative to their high-luminosity cousins, and (ii) because their low gamma-ray flux make them difficult to detect with conventional electromagnetic detectors (e.g. {\it Swift}). 
Recently, Murase \& Ioka~\cite{Murase:2013ffa} showed that choked jets may be more favorable as 
sites of efficient neutrino production. Jets which successfully penetrate both the progenitor star -- 
and if applicable a circumstellar envelope -- (i.e. emergent jets) typically have high-luminosities such
that they form radiation-mediated shocks, which are unfavorable for CR acceleration and neutrino 
production. Taking into account the luminosity and redshift distribution of LL GRBs, we show that 
they and the choked jets may contribute to the diffuse neutrino flux while remaining absent from 
GRB joint electromagnetic-neutrino searches. We also explicitly show the conditions required to produce choked jets with radiation-unmediated shocks.

\begin{figure*}[tb]
\label{fig:setup}
\includegraphics[width=0.32\linewidth]{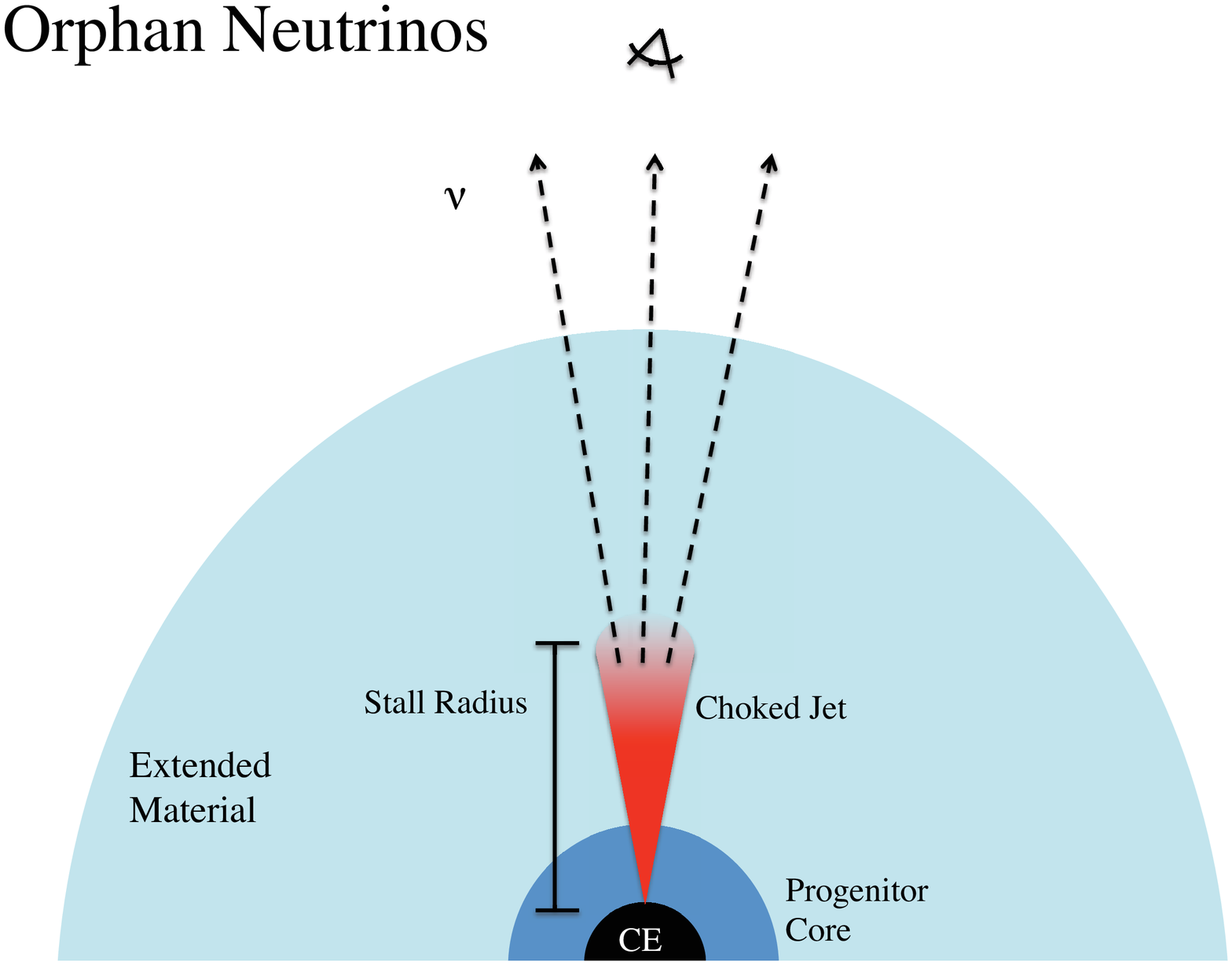}\hfill
\includegraphics[width=0.32\linewidth]{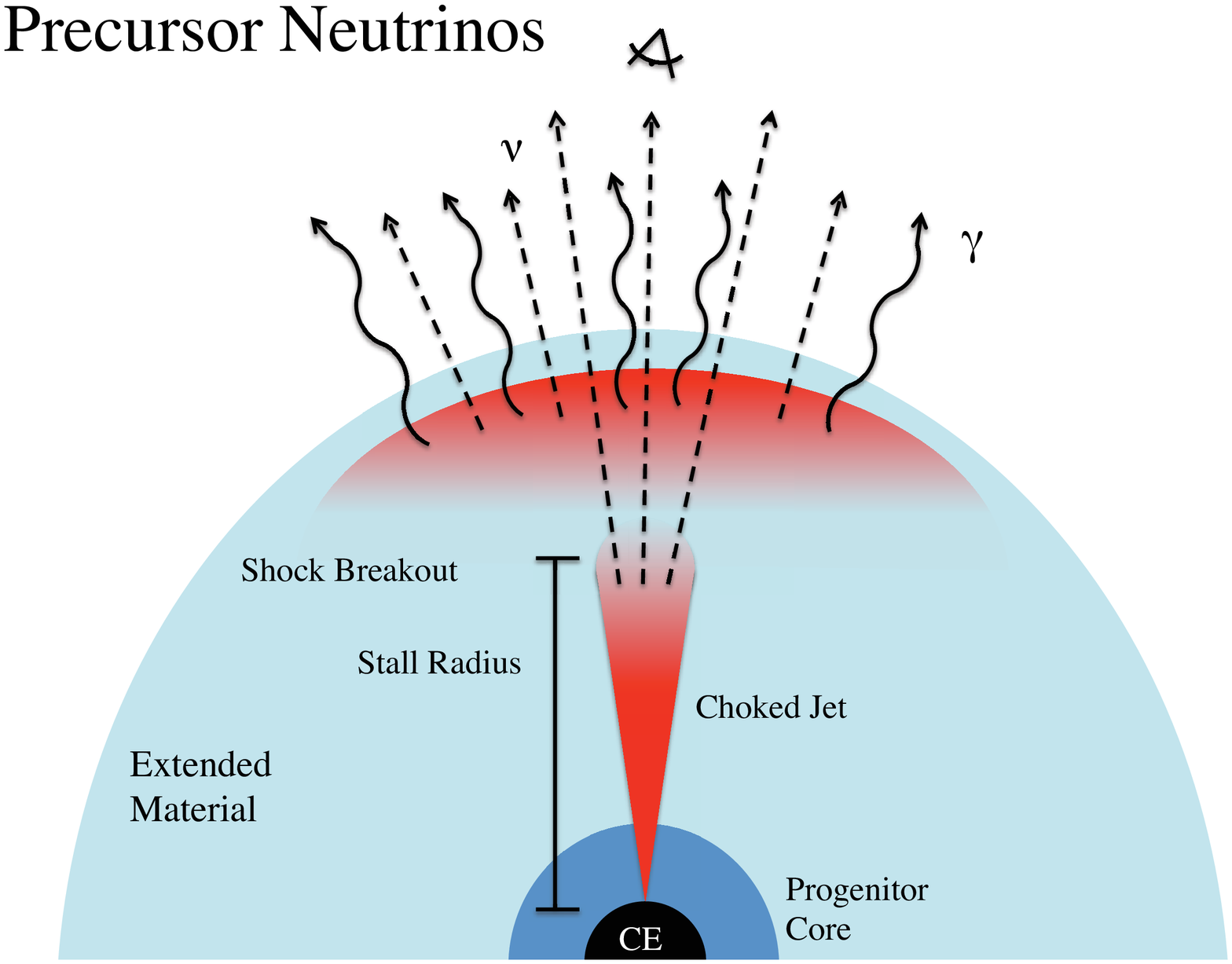}\hfill
\includegraphics[width=0.32\linewidth]{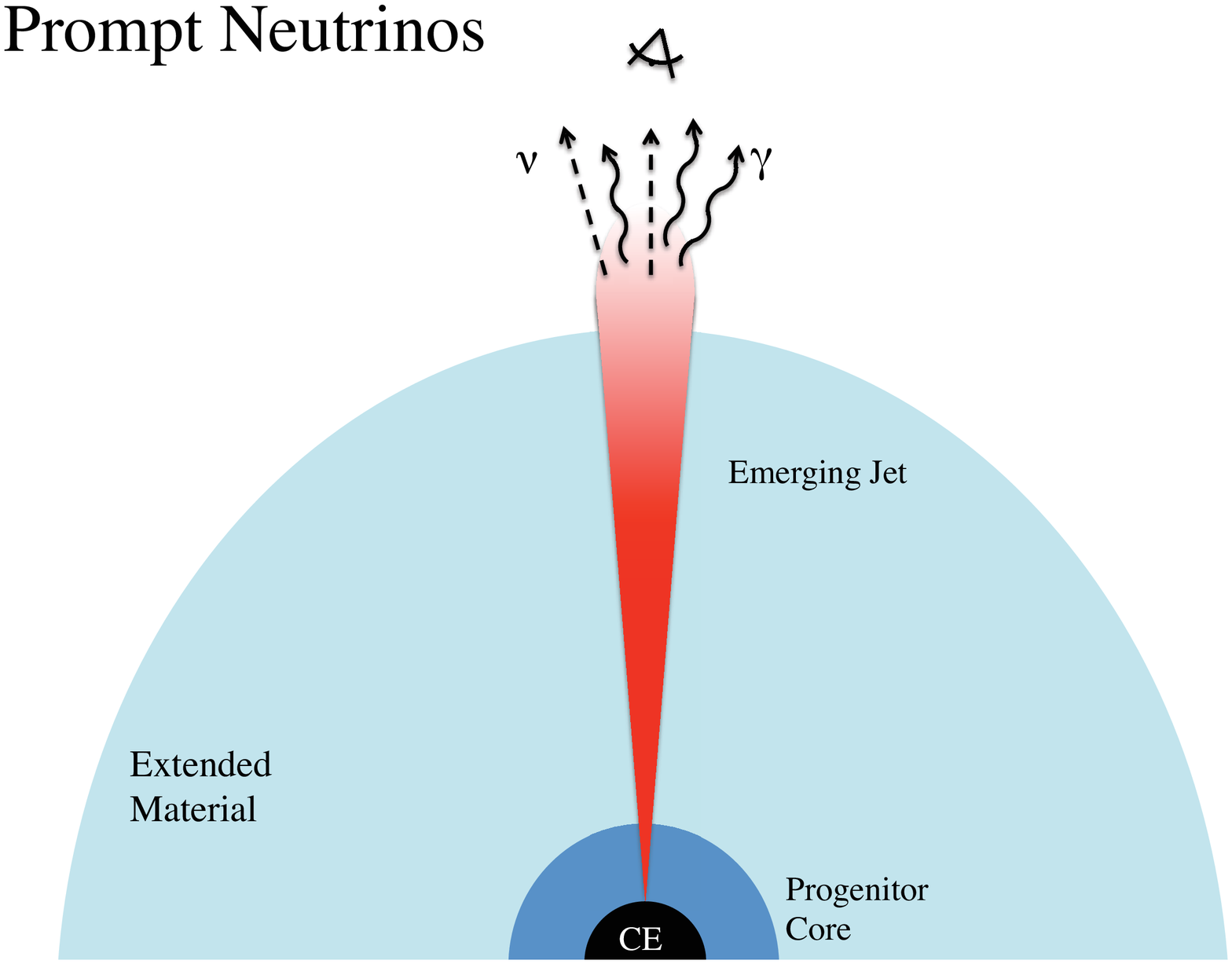}
\caption{{\bf Left panel:} The choked jet model for jet-driven SNe.  Orphan neutrinos are 
expected since electromagnetic emission from the jet is hidden, and such objects may be 
observed as hypernovae. 
{\bf Middle panel:} The shock breakout model for LL GRBs, where transrelativistic shocks are 
driven by choked jets. A precursor neutrino signal is expected since the gamma-ray 
emission from the shock breakout occurs significantly after the jet stalls (e.g.,~\cite{Meszaros:2001ms}).
{\bf Right panel:} The emerging jet model for GRBs and LL GRBs.  Both neutrinos and 
gamma-rays are produced by the successful jet, and both messengers can be observed as 
prompt emission. 
\label{fig1}
}
\end{figure*}

\section{Dynamics of Relativistic Jets} 
\label{sec:jetprop}
\subsection{Model Setup for Emergent Jet, Shock Breakout, andChoked Jet Scenarios} 
\label{sec:pictures}
GRBs are thought to result from the intense emission from relativistic jets that successfully penetrate a progenitor star, and an understanding of jet propagation is undoubtedly relevant~\citep[e.g.,][]{Meszaros:2001ms,Bromberg:2011fg,Mizuta:2013yma}. 
It would be natural to expect that the radiation mechanism of LL GRB gamma-ray emission is similar to that of classical GRBs \cite{Toma:2006iu,irwin+chevalier15jets,ghisellini+07-060218}.  The simplest such model is a scaled-down version of the classical GRB, where dissipation occurs in a mildly relativistic jet which has emerged outside of the progenitor star  and any circumstellar material. We call this scenario the emerging jet (EJ) model (see Fig.~1 right panel). For EJs, prompt neutrino emission is produced together with prompt gamma-ray emission outside the star, identical to the scenario expected from classical GRBs~\cite{Murase:2006mm,Gupta:2006jm,Liu:2011cua}. 

Another interpretation of LL GRBs which has received attention is the shock breakout emission model, where the prompt emission is attributed to dissipation caused by a transrelativistic, aspherical shock in a dense wind ~\cite{campana+06-060218,waxman+07sb,Bromberg:2011fm,Nakar:2011mq}.
The origin of the relativistic velocity components in the ejecta is an issue. One of the promising possibilities is that the fast shock is driven by a choked jet.  The jet stalls close enough to the photosphere so that a transrelativistic shock breaks out through the star and its extended material.  We call this model the choked jet-shock breakout (CJ-SB) model, and the middle panel of Fig.~1 shows its schematic. 

A luminous jet naturally leads to an easier break out from a typical compact progenitor such as a 
Wolf-Rayet (WR) star. One possible cause for an energetic jet to stall is through smothering 
by an extended envelope of wind material. Such an environment need be only marginally more massive 
than what is inferred, e.g. by \cite{campana+06-060218,waxman+07sb}. For an extended material mass 
$M_{\rm ext}\sim{10}^{-3}-{10}^{-2}M_\odot$, a typical GRB jet penetrating the compact star is expected to be choked at a distance of $\sim{10}^{12}-{10}^{13}$~cm from the central engine~\cite{nakar15-060218}. Although such a radius is 10 times larger than that of WR stars believed to be GRB progenitors, it is appealing that the model can also explain the mysterious UV component in GRB 060218. Envelopes of $M_{\rm ext} \sim 10^{-7}~M_\odot$ are observed for pre-exploded WRs, but there is increasing evidence for exceptionally high mass loss in the weeks prior to collapse~\cite{ofek+10wind,smith+11ccsne}. Accompanying theoretical explanations have been proposed~\cite{shiode+quataert14wind,Chevalier:2012ba,moriya15wind}.

Choked jets with a shock breakout component  may or may not produce prompt gamma-rays ~\citep[e.g.,][]{waxman+07sb,wang+07sb}.
The jet may stall sufficiently far below the photosphere that the piston action of the jet does not lead to transrelativistic velocity components of the shock, i.e., such objects are simply observed as energetic SNe or hypernovae in the optical band with no accompanying gamma-ray emission.  We call this final scenario the choked jet (CJ) model. That is, the SN has a ``normal" quasispherical shock, without a jet-driven (aspherical, transrelatvistic) extra shock breakout component (see Fig.~1 left panel where we have neglected to show the subrelativistic shock for clarity).

Thus, it is possible to have a unified picture where the GRB-SN connection is explained by the strength of the choked jets. Both of the CJ and CJ-SB models provide favorable environments for neutrino production since shock acceleration may occur inside the jet before the shocks become radiation mediated, allowing for efficient neutrino emission.
On the other hand, the gamma-rays produced deep inside the choked jets will not be able to escape through the extended material to observers, and bright gamma-rays are observed from the shock breakout component only.  
A similar picture for neutrino production is considered in Ref.~\cite{Murase:2013ffa} for low-power jets with $L\lesssim{10}^{47}~{\rm erg}~{\rm s}^{-1}$ and/or extended progenitors such as blue supergiants (BSGs).

\subsection{Hydrodynamical Constraints on Choked Jets} 
\label{sec:stall}
We will consider VHE neutrino emission from choked jets and it is relevant to consider the condition where relativistic jets are stalled. The dynamics of a relativistic jet is determined by its interaction with the ambient medium in the progenitor star and circumstellar envelope, which can change the shape of the jet through collimation shocks \citep{Bromberg:2011fg,Mizuta:2013yma}. While this work focuses on neutrino emission only from internal and termination shocks, it is important to note that collimation shocks near the base of the jet will also affect estimates of VHE neutrino production~\cite{Murase:2013ffa} but are generally not considered in most of the previous literature~\cite{Razzaque:2003uv,Razzaque:2004yv,Ando:2005xi,Horiuchi:2007xi,2015arXiv151201559T}.
As the jet drills through the star, a contact discontinuity is formed between the shocked jet material and 
the shocked ambient matter. This region of shocked material is often referred to as the jet head. Balancing the jet's internal pressure with the ram pressure of the ambient material determines the head's dynamics (see, e.g.,~\cite{Meszaros:2001ms,Bromberg:2011fg,Mizuta:2013yma}), and the head velocity is given by
\bea
\label{eq:head}
\beta_h=\frac{\beta_j}{1+\tilde{L}^{-1/2}},
\ena
with dimensionless luminosity 
\bea
\tilde{L} \approx \frac{L_{0j}}{\pi(r_h\theta_0)^2\rho_ac^3},
\ena
where $L_{0j}$ is the one-sided jet luminosity, $\rho_a$ is the ambient density, $r_h$ is the distance of the jet head from the central engine, and $\theta_0$ is the initial opening angle inside the star.  Assuming the jet material is relativistic $\beta_j\sim1$ before reaching the head it is obvious from Eq. (\ref{eq:head}) that for $\tilde{L}\gg1$ the head moves relativistically as well. Then, the jet will be collimated for $\tilde{L}\ll \theta_0^{-4/3}$ or uncollimated for $\tilde{L}\gg\theta_0^{-4/3}$ \cite{Bromberg:2011fg}. 

First, let us consider a jet propagating inside its progenitor star. As shown in Refs.~\cite{Bromberg:2011fg,Mizuta:2013yma}, such a jet is typically collimated. Let us assume that the density profile is approximated to be $\rho_a=(3-\alpha)M_*{(r/R_*)}^{-\alpha}/(4\pi R_*^3)$ ($\alpha\sim1.5-3$). Here $M_*$ is the progenitor mass and $R_*\sim0.6-3R_{\odot}$.  For WR progenitors, we may take $\alpha=2.5$ \cite{2006ApJ...637..914W}, leading to the jet head radius $r_h\simeq5.4\times{10}^{10}~{\rm cm}~t_{1}^{6/5}L_{0,52}^{2/5}{(\theta_0/0.2)}^{-4/5}{(M_{*}/20~M_{\odot})}^{-2/5}\\R_{*,11}^{1/5}$, where $L_0=4L_{0j}/\theta_0^2$ is the isotropic-equivalent total jet luminosity~\cite{Bromberg:2011fg,Murase:2013ffa}. The classical GRB jet is typically successful (i.e., it emerges from the progenitor), since the time required for the jet to escape the progenitor $t_{\rm jbo}\approx17~{\rm s}~L_{0,52}^{-1/3}{(\theta_0/0.2)}^{2/3}{(M_{*}/20~M_{\odot})}^{1/3}R_{*,11}^{2/3}$ is shorter than the jet duration $t_{\rm eng}\sim{10}^{1.5}$~s.  This time is in good agreement (i.e. within a factor of a few) with numerical studies of jet emergence \cite{2003ApJ...586..356Z, 2007ApJ...665..569M, 2009ApJ...699.1261M, Bromberg:2011fg}. See also Fig.~15 of Ref.~\cite{Mizuta:2013yma}.

Toma et al. \citep{Toma:2006iu} suggested that the prompt emission of GRB 060218 may come from an emerging jet with a Lorentz factor of $\Gamma\sim5$, and this possibility of marginally successful jets has been further investigated by Irwin \& Chevalier \cite{irwin+chevalier15jets}.  The jet has more difficulty in penetrating the progenitor star due to its lower luminosity, but on the other hand, its longer duration helps in achieving breakout.  In this model, the prompt gamma-ray emission may come both from relatively low radii around the photosphere or large radii.  
Such marginally successful jets are expected for larger radius progenitors such as BSGs, and UL GRBs may correspond to the case of successful GRBs~\cite{Murase:2013ffa}.  

Next, we consider jets embedded in an extended, massive envelope. The jet can be choked if the mass of the extended material is sufficiently large. Motivated by the CJ-SB model for LL GRBs, we consider an extended material with mass $M_{\rm ext}\sim{10}^{-2}~M_\odot$ and radius $r_{\rm ext}\sim3\times{10}^{13}$~cm.  WR stars have been observed with such unusually massive envelopes in the months leading up to their SN explosion \cite{2013Natur.494...65O, 2014Natur.509..471G, 2014ApJ...789..104O}. Nakar \cite{nakar15-060218} suggested similar envelope parameters for LL GRB 060218, but without strong constraints on the density profile. For simplicity, we therefore assume the same wind profile for all LL GRBs, namely 
\beq
\rho(r)=5.0\times{10}^{-11}~{\rm g}~{\rm cm}^{-3}~\left(\frac{M_{\rm ext}}{0.01~M_\odot}\right)r_{\rm ext,13.5}^{-3}{\left(\frac{r}{r_{\rm ext}}\right)}^{-2},
\enq
with the density at the outer envelope edge $\rho_{\rm ext}\equiv\rho(r_{\rm ext})/(5.0\times{10}^{-11}~{\rm g}~{\rm cm}^{-3})$. Assuming -- as Nakar did -- that the majority of the envelope's mass is located near the outer radius (i.e. the quantity $\rho(r)\, r^3$ increases up until $r_{ext}$) different wind profiles do not significantly affect the dynamics of the jet head. The jet is typically uncollimated for sufficiently luminous jets and the Lorentz factor of the jet head is given by
\beq
\Gamma_h\approx\frac{\tilde{L}^{1/4}}{\sqrt{2}}\simeq3.5~L_{0,52}^{1/4}\rho_{\rm ext}^{-1/4}r_{h,13.5}^{-1/2},
\enq
while the jet head radius is estimated to be
\beq
r_h\approx2\Gamma_h^2 ct\simeq2.3\times{10}^{13}~{\rm cm}~L_{0,52}^{1/2}\rho_{\rm ext}^{-1/2}r_{\rm ext,13.5}^{-1}t_{1.5}.
\enq
The condition $r_h=r_{\rm ext}$ gives the jet breakout time $t_{\rm jbo,ext}$, and the condition $t_{\rm jbo,ext}\lesssim t_{\rm eng}$ gives the jet-stalling condition
\bea
\label{eq:js}
L_\gamma\lesssim L_{\gamma}^{\rm JS}&\approx&0.95\times{10}^{48}~{\rm erg}~{\rm s}^{-1}~\left(\frac{\epsilon_\gamma}{0.25}\right){\left(\frac{\theta_j}{0.2}\right)}^2t_{\rm eng,1.5}^{-1}\nonumber\\
&\times&T_{3.5}^{-1}\rho_{\rm ext} r_{\rm ext,13.5}^4,
\ena
where we have used
\beq
L_{\gamma}\approx \epsilon_\gamma \frac{\theta_j^2}{2}\frac{L_0t_{\rm eng}}{T}. 
\enq
$L_{\gamma}$ is the observed luminosity of the LL GRB, $\epsilon_\gamma$ is the gamma-ray  emission efficiency, $\theta_j$ is the choked jet opening angle in the extended material, and $T$ is the typical observed duration of LL GRBs. For choked jets, the jet head radius at $t_{\rm eng}$ is defined as the jet-stalling radius $r_{\rm stall}$. In our CJ-SB scenario, the central engine activity time $t_{\rm eng}$ is unrelated to the duration of the prompt emission $T$, since the later only depends the shock velocity and breakout radius $T\approx r_{\rm sb}/\Gamma_{\rm sb}^2c$, which are determined from the envelope properties. The former timescale is deduced from the lifetime of GRB jets that are seen in high-luminosity GRBs, while the later reflects the typical observed duration of a LL GRB.  Note that the hydrodynamic constraints are relevant for neutrino production. First, they restrict the emission radius, which limits the overall non-thermal particle energy density as well as the maximum neutrino energy. Since the emission region of traditional GRB jets assume a wide variety of values (e.g. $10^{11}~{\rm cm} \lesssim r_{\rm em} \lesssim 10^{17} $ cm for the classic fireball model in \cite{Bustamante:2014oka}), the phenomenology of neutrinos from choked jets can be quite different.  Additionally, the jet luminosity and central engine duration need to be consistently determined. If the jet is too powerful or its duration is too long, it is no longer choked and should be reduced to the classical GRB case.

If the jet is choked in the dense wind close to the edge of the star, it will launch a transrelativistic shock that becomes an aspherical shock breakout.  As described in Refs.~\cite{Katz:2011zx,Kashiyama:2012zn}, breakout nonthermal emission may be released when the optical depth of the shock reaches unity.  The emission time of the breakout -- and therefore the approximate duration of the subsequent GRB -- is $T\approx r_{\rm sb}/(\Gamma_{\rm sb}^2c)\sim{10}^{3.5}~{\rm s}$ in agreement with the average duration of LL GRBs, where $r_{\rm sb}$ is the shock breakout radius and $\Gamma_{\rm sb}$ is the Lorentz factor of the shock.  It has been shown that shock breakouts produce smooth light curves similar to what are seen in LL GRBs.

\section{Radiation Constraints on Shock Acceleration}
CRs are generally assumed to be accelerated with a power-law distribution by the first-order Fermi process in the presence of shocks or turbulence. As known from the literature of nonrelativistic shocks, (e.g.,~\cite{Murase:2010cu,Katz:2011zx}), efficient conversion of the fluid kinetic energy to a non-thermal particle population can occur if the shocks are collisionless (i.e. mediated by plasma instabilities), requiring the upstream plasma to be optically thin for relativistic shocks. CRs gain energy thanks to the shock compression. If the shock is mediated by radiation, efficient acceleration is prevented~\cite{Levinson:2007rj,Murase:2013ffa}
since the shock width is larger than the CR Larmor radii and particles cannot efficiently cross between the upstream and downstream fluids. This subltle feature of CR acceleration in relativistic jets is often not considered in the literature~\cite{Razzaque:2003uv,Razzaque:2004yv,Ando:2005xi,Horiuchi:2007xi,2015arXiv151201559T}.

The radiation constraints give us stringent restrictions on the rate VHE neutrino production from choked jets. Two shells in the jet have a relative Lorentz factor $\Gamma_{\rm rel}\approx \Gamma_r/2\Gamma$, where the rapid shell is moving with velocity $\Gamma_r \sim {\rm few}\times \Gamma$.  Murase \& Ioka~\cite{Murase:2013ffa} derived radiation constraints.  For internal shocks, efficient CR acceleration can occur if $\tau_T=n'_{j}\sigma_T(r_{\rm is}/\Gamma)\lesssim{\rm min}[\Gamma_{\rm rel}^2,0.1C^{-1}\Gamma_{\rm rel}^3]$ or
\begin{equation}
L_{52}r_{\rm is,10}^{-1}\Gamma_2^{-3}\lesssim5.7\times{10}^{-3}~{\rm min}[\Gamma_{\rm rel,0.5}^2,0.32C_{1}^{-1}\Gamma_{\rm rel,0.5}^3],
\end{equation}
where $\Gamma\sim100$ is the bulk Lorentz factor of the jet, $L$ is the isotropic-equivalent kinetic luminosity of the jet, and the comoving density is $n'_{j}\approx L/(4\pi\Gamma^2r_{\rm is}^2m_pc^3)$. The factor $C = 1 + 2\ln\Gamma_{rel}^2$ accounts for possible pair production effects in the shocked material \cite{Nakar:2011mq,2010ApJ...725...63B}.
The jet propagating in the star is usually collimated, and the collimation shock radius $r_{\rm cs}$ is smaller than the jet head radius ($r_{\rm cs}<r_{h}<R_*$). We expect radiation-mediated shocks for the jet inside a WR star, but the shocks can be collisionless if the ambient material is more extended~\cite{Murase:2013ffa}.  Setting the shock radius equal the stall radius calculated in the previous section ($r_{\rm is}=r_{\rm stall}$) and ignoring the prefactor depending on details of the shock structure, we obtain the following upper limit for the luminosity based on radiation constraints,
\begin{equation}
L\lesssim1.7\times{10}^{54}~{\rm erg}~{\rm s}^{-1}~\Gamma_2^6t_{\rm eng,1.5}^{2}\rho_{\rm ext}^{-1}r_{{\rm ext},13.5}^{-2}.
\end{equation}  
In the CJ-SB model, using the observed gamma-ray luminosity, the constraint becomes
\begin{eqnarray}
\label{eq:rc}
L_{\gamma}\lesssim L_{\gamma}^{\rm RC}&\approx&8.6\times{10}^{49}~{\rm erg}~{\rm s}^{-1}~
\left(\frac{\epsilon_\gamma}{0.25}\right){\left(\frac{\theta_j}{0.2}\right)}^2\nonumber\\
&\times&\Gamma_2^6t_{\rm eng,1.5}^{3}{T}_{3.5}^{-1}\rho_{\rm ext}^{-1}r_{{\rm ext},13.5}^{-2}.
\end{eqnarray}  
As the luminosity increases or the radius decreases, the optical depth largely exceeds unity and CRs cannot be accelerated efficiently at the shock. While we assume the plasma is optically thin inside the jet, the envelope or circumstellar medium outside is largely optically thick to Thomson scattering. Therefore, photons are free to move inside the jet but cannot escape. The free streaming of photons from outside the jet core -- specifically thermal emission from the jet head -- allow for efficient neutrino production through $p\gamma$ interactions. 

The radiation constraints apply to shocks in the envelope material as well as those in the choked jet.  Before breakout, the shock is radiation mediated. As photons diffuse out from the system, the shock becomes collisionless and CRs may be accelerated~\cite{Katz:2011zx,Kashiyama:2012zn}.  The condition is given by $\tau_T\lesssim \beta_{\rm SH}^{-1}$, where $\beta_{\rm SH}=V_{\rm SH}/c$ is the shock velocity.

\section{Neutrino Production in Choked Jets} 
\label{sec:nuprod}
In the present work, we focus on the CJ and CJ-SB models, so we assume that CRs are accelerated in jets choked by the circumstellar material. For simplicity we assume that internal shocks occur near the collimation shock radius since they are expected to occur most frequently at $r_{\rm is}\approx2\Gamma_j^2c\delta t$ when the jet is not stalled, where $\delta t$ is the variability timescale.  
We calculate neutrino spectra only for GRB jets satisfying the following condition, 
\beq
L_{\gamma}\lesssim {\rm min}[L_{\gamma}^{\rm JS},L_{\gamma}^{\rm RC}],
\enq
where $L_{\gamma}^{\rm JS}$ comes from the jet-stalling condition (Eq. \ref{eq:js}) and $L_{\gamma}^{\rm RC}$ comes from the radiation constraint (Eq. \ref{eq:rc}).  The neutrino spectra are calculated numerically, taking into account various microphysical processes such as multipion production.  For details of the method, see Refs.~\cite{Murase:2005hy,Murase:2007yt,Murase:2008sp,Murase:2013ffa}.  
We explicitly calculate the CJ component of the neutrino flux from jets with luminosities $L_{\gamma}={10}^{45}~{\rm erg}~{\rm s}^{-1}$, $L_{\gamma}={10}^{46}~{\rm erg}~{\rm s}^{-1}$, $L_{\gamma}={10}^{47}~{\rm erg}~{\rm s}^{-1}$, and $L_{\gamma}={10}^{48}~{\rm erg}~{\rm s}^{-1}$ and use the results to infer the approximate contribution from CJs with other luminosities in Eq. \ref{eq:difflux} below.
For the GRB jet parameters, assuming that the jets leaving the star are similar to those of classical GRBs, we use $t_{\rm eng}={10}^{1.5}$~s, $\theta_j=0.2$, and $\Gamma=100$. 
In this work, we calculate the neutrino emission for different luminosities but fix the other parameters because of computational limitations. Note that our GRB jet parameters have been widely used in the GRB literature as typical values. Also, thanks to the high neutrino production efficiency, the flux level is insensitive to the GRB jet parameters~\cite{Murase:2008sp,Murase:2013ffa}. The resulting neutrino flux effectively scales with the CR loading parameter.
Note that lower Lorentz factors further restrict the parameter space of the VHE neutrino production since the shock becomes radiation mediated. For slow jets with $\Gamma\lesssim10$, VHE neutrino production inside a typical GRB progenitor requires $L\lesssim{10}^{47}~{\rm erg}~{\rm s}^{-1}$~\cite{Murase:2013ffa}. In addition, we assume $M_{\rm ext}=0.01~M_\odot$ and $r_{\rm ext}=10^{13.5}$~cm for the extended material, which are based on the parameters suggested for the explanation of LL GRBs.

CRs accelerated at internal shocks interact with photons produced from both internal shocks and the jet head. The properties of electrons accelerated at internal shocks are uncertain so we conservatively consider only thermal photons coming from the jet head~\cite{Meszaros:2001ms}.  In rest frame of the head the photon temperature is $kT_h\simeq0.37~{\rm keV}~L^{1/4}_{52}r^{-1/2}_{h,13}\Gamma^{-1/4}_{h,0.5}$. 
This temperature is Lorentz boosted by a factor $\Gamma_{\rm rel} \sim \Gamma/2\,\Gamma_h$ in the jet rest frame. Likewise, the photon energy density seen in the jet is $U_{\gamma,j} \sim \Gamma_{\rm rel}^2 U_{\gamma,h}$, which can make the thermal component from the head the most significant photon field. We also take into account the photon escape fraction $\sim{(n_{\gamma,h}\sigma_Tr_h/\Gamma_h)}^{-1}$.  The photomeson production efficiency satisfies
\beq
{\rm min}[1,f_{p\gamma}]\approx1
\enq
Thus, CRs that exceed the pion production threshold are depleted by the photomeson production. Choked jets can be regarded as ``calorimetric'' sources in the sense that all of the available CR energy goes into making neutrinos and the observation of neutrinos allow us to directly probe the amount of accelerated CRs.  Note that, although there are nonthermal populations of photons radiated by co-accelerated pairs, this point is unchanged.  Additional photons enhance the efficiency of the photomeson production.

At subphotospheric radii, inelastic $pp$ interactions are shown to be relevant below 100~TeV and the photon meson efficiency is found to dip due to the Bethe-Heitler process ~\cite{Murase:2008sp,Gao:2012ay}. When radiation constraints are satisfied, the $pp$ optical depth during the dynamical time in the jet is limited to~\cite{Murase:2008sp,Kashiyama:2012zn}
\beq
f_{pp}\lesssim\frac{\kappa_{pp}\sigma_{pp}}{\sigma_T}\simeq0.04.
\enq
Thus, for the CR spectrum with $s\sim2$, the energy flux of the $pp$ component is typically lower than that of the associated $p\gamma$ component, and the main production mechanism for VHE neutrinos inside the jet is the photomeson production process. 
Note that, as pointed out in Ref.~\cite{Murase:2013ffa}, low-energy CRs in the jet may eventually be advected along the collimated jet and all CRs can be depleted for neutrino production via subsequent $pp$ interactions. However, for uncollimated shocks, CRs accelerated at the shocks may simply cool via adiabatic losses during the jet expansion. In this work, to be conservative, we do not consider effects of the remaining CRs. 

\begin{figure}[t]
\includegraphics[width=3.5in]{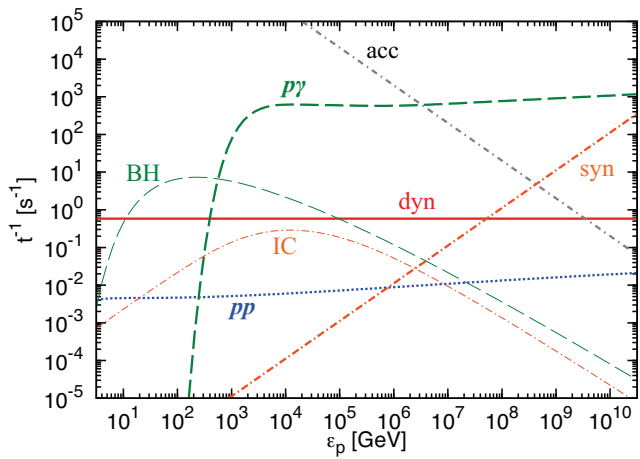}
\caption{Various energy-loss rates of CR protons in the CJ model for LL GRBs as a function of the comoving proton energy $\varepsilon$. Photomeson production ($p\gamma$), Bethe-Heitler pair production (BH), hadronuclear ($pp$), synchrotron radiation (syn), inverse-Compton radiation (IC), adiabatic cooling (dyn), and acceleration (acc) processes are considered.  The case of $L_{\gamma}={10}^{47}~{\rm erg}~{\rm s}^{-1}$ (implying $L=2\times{10}^{51}~{\rm erg}~{\rm s}^{-1}$) is shown.  
}
\vspace{-1.\baselineskip}
\label{fig:cooling} 
\end{figure}

CRs lose their energy via photomeson production ($p\gamma$), Bethe-Heitler pair production (BH), hadronuclear ($pp$), synchrotron radiation (syn) and inverse-Compton radiation (IC) processes. Since in our case the jet is expanding, adiabatic losses are also included.  
The numerically calculated acceleration and cooling timescales are shown in Fig.~\ref{fig:cooling}.  At higher energies, photomeson production is the dominant cooling channel for CRs, and sets a maximum CR energy of $\varepsilon_p^{M}\sim 3$~PeV in the comoving frame. At lower energies $\varepsilon_p\lesssim1$ TeV, BH cooling dominates and suppresses the neutrino spectrum around 100~GeV.  The magnetic field is set by using the parameter $\epsilon_B=L_B/L_0=0.1$, and the acceleration time is given by $t_{\rm acc}=\varepsilon_p/(eBc)$. 
Note that, even if the initial maximum CR energy is very high, the energy of escaping CRs is low since they are depleted advecting to the shock downstream. 

The neutrino products receive $\sim0.05\%$ of their parent CR energy. For the energies of interest, $E_\nu\sim30$~TeV, the $\Delta$ resonance in $p\gamma$ interactions requires target photons with energy $\sim0.2~{\rm keV}-0.2~{\rm MeV}$ depending on the Lorentz factor of the target photon field (i.e. between the jet interior and head) ~\cite{Murase:2015xka}. The jet head appears as a blackbody with temperature $kT_{h}\sim0.1-1$~keV as seen from the jet core providing photons within the correct energy range. 
The fraction of CRs that interact with these photons using the box approximation of Ref.~\cite{Waxman:1997ti} is $f_{p\gamma} \gg 1$, which is also seen from Fig.~\ref{fig:cooling}.  As expected, $p\gamma$ interactions dominate the neutrino production. 

In addition, meson cooling is also taken into account by solving the kinetic equations numerically~\citep{Murase:2005hy,Murase:2007yt,Murase:2013ffa}. When the jet-stalling condition and radiation constraints are considered, we find that the pion and muon components are almost always dominant and the kaon component could be relevant only above PeV energies.

\section{Diffuse Neutrinos from Low-Luminosity GRBs and Hypernovae}
Finally, we calculate the diffuse neutrino flux by convolving the neutrino spectra for different luminosities with $10^{45}~{\rm erg}~{\rm s}^{-1}\lesssim L_\gamma \lesssim{10}^{48}~{\rm erg}~{\rm s}^{-1}$. The upper luminosity limit is found by constraining the jet to be choked with shocks that are not radiation dominated. The lower luminosity limit is chosen such that the results are not sensitive to this choice.
As emphasized above, contrary to predictions for neutrino emission from optically-thin environments, we do not have much uncertainty in values of $f_{p\gamma}$, i.e., the $p\gamma$ efficiency is always close to the maximum. Thus, as long as the rate uncertainty is not too large, the only critical parameter is the total energy of CRs even though there are other subparameters such as $M_{\rm ext}$ and $r_{\rm ext}$. In this work, the jet kinetic energy is assumed to be similar to that of classical GRBs.  Here, importantly, even if the observed GRB luminosity is low (recall that ``low-luminosity GRBs'' here are defined based on the observed luminosity), choked jets themselves may be as powerful as the jets of classical high-luminosity GRBs. In the CJ-SB model, the choked jet has isotropic-equivalent luminosity $L\sim{10}^{51}-{10}^{52}~{\rm erg}~{\rm s}^{-1}$, but the observed gamma-ray luminosity is smaller by a factor of $(2/\theta_j^2)(T/t_{\rm eng})$. (Clearly, $t_{\rm eng}$ can also play a large role in determining whether a jet will give rise to a classical GRB or an LL GRB).  For the shock breakout luminosity $L_\gamma$, the total absolute CR energy in the jet is assumed to be ${\mathcal E}_{\rm CR}=(\epsilon_{\rm CR}/\epsilon_\gamma)(L_\gamma T)\simeq6.3\times{10}^{50}~{\rm erg}~(\xi_{\rm CR}/2)L_{\gamma,47}T_{3.5}$ (where $\xi_{\rm CR}\equiv\epsilon_{\rm CR}/\epsilon_\gamma=2(0.25/\epsilon_\gamma)(\epsilon_{\rm CR}/0.5)$ is the so-called CR loading factor).  Note that the total absolute CR energy scales as the observed gamma-ray luminosity. Also, the CR spectrum is assumed to be $dN_p/d{\varepsilon'}_p\propto {\varepsilon'}_p^{-2}$. 
 
The diffuse neutrino flux is calculated via~(e.g.,~\cite{Murase:2007yt})
\begin{eqnarray}
\label{eq:difflux} 
\Phi_\nu &=& \frac{c}{4\pi H_0}\int_{z_{\rm min}}^{z_{\rm max}}\,dz \,\int_{L_{\rm min}}^{L_{\rm max}}\,dL_{\gamma}\nonumber\\
&\times&\frac{dR_{\rm cho}(z)/dL_\gamma}{\sqrt{\Omega_M\,(1+z)^3 + \Omega_\Lambda}}\,\left(\frac{dN_{\nu}((1+z)E_\nu)}{dE'_\nu}\right),
\end{eqnarray}
where $dN_\nu/dE'_\nu$ is the neutrino spectrum per burst, $H_0$ is the Hubble constant, $\Omega_M$ and $\Omega_\Lambda$ are cosmological parameters. If LL GRB progenitors evolve as the star-formation rate (SFR), we rescale the function found by~\cite{yuksel+08sfr} 
\begin{eqnarray}
\label{eq:SFR}
R_{\rm cho}(z)&=&f_{\rm cho}R_{\rm LL}\nonumber\\
&\times&\left[(1+z)^{p_1\kappa} + \left(\frac{1+z}{5000}\right)^{p_2\kappa}
+\left(\frac{1+z}{9}\right)^{p_3\kappa}\right]^{1/\kappa},
\end{eqnarray}
with $\kappa = -10$, $p_1 = 3.4$, $p_2 = -0.3$, $p_3 = -3.5$, $f_{\rm cho}$ expresses the contribution of choked jets without shock breakout (i.e., orphan neutrinos), and $R_{\rm LL}\sim100-200~{\rm Gpc}^{-3}~{\rm yr}^{-1}$ is the local LL GRB rate at $z=0$.
Ref.~\cite{sun+15rates} constructed a luminosity function (i.e., the number of bursts with an observed isotropic-equivalent luminosity within a given luminosity interval) uniquely for the LL GRB population
\begin{equation}
\label{eq:LF}
\frac{d R_{\rm LL}}{dL_\gamma}\approx\frac{(\alpha - 1)R_{\rm LL}}{L_m}\,\left(\frac{L_\gamma}{L_m}\right)^{-\alpha}.
\end{equation}
It was found that the data was fit best with a local rate of $R_{\rm LL}=164^{+98}_{-65}\,\mathrm{Gpc^{-3}\,yr^{-1}}$, index $\alpha = 2.3\pm 0.2$ and characteristic luminosity $L_m = 5\times10^{46}\,\mathrm{erg \, s^{-1}}$. 

\begin{figure}[t]
\includegraphics[width=3.6in]{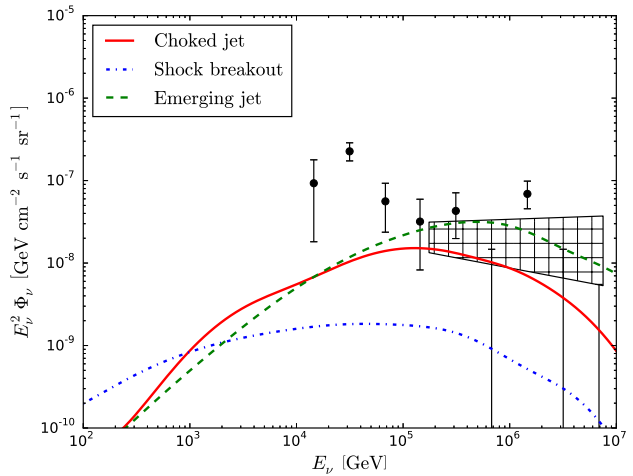}
\caption{All-flavor diffuse VHE neutrino fluxes from LL GRBs in various models. The choked jet CJ (this work), shock breakout (CJ-SB) ~\cite{Kashiyama:2012zn}, and emergent jet (EJ)~\cite{Murase:2006mm} components are shown. The shock breakout component has been updated to include the newest luminosity function and redshift evolution, while the EJ component is luminosity insensitive with the redshift evolution of Ref.~\cite{Porciani:2001kh} and is shown for illustrative purposes. Note that neutrinos are observed as prompt emission or precursor emission. The IceCube data based on the combined analysis \cite{Aartsen:2015ita} and up-going muon neutrino analysis~\cite{Ishihara:2015tev} are overlaid.  
}
\vspace{-1.\baselineskip}
\label{fig:flux} 
\end{figure}

Fig. \ref{fig:flux} shows the diffuse neutrino flux from LL GRBs for different components.  
For our parameter set in the CJ-SB model that explains LL GRBs, we find that the diffuse neutrino flux is compatible with the measured flux for $E_{\nu}\sim0.1-1$~PeV.  
There are three relevant remarks. 
(i) First, since the gamma-rays and the dominant component of neutrinos are produced in different regions, a prediction of the CJ-SB model is that the majority of the LL GRB neutrino signal arrives $(r_{\rm sb} - r_{\rm stall})/c \sim 100-1000$~s before the LL GRB triggers a detector. 
(ii) Second, the VHE neutrino emission from choked jets is highly beamed in the CJ-SB model. On the other hand, the shock breakout contribution is nearly isotropic so that the associated neutrino emission can be observed from off-axis observers~\cite{Kashiyama:2012zn}.  
(iii) Third, precursor neutrinos from choked jets will be found within a much smaller temporal window ($t_{\rm eng}\sim{10}^{1.5}$~s) compared to the electromagnetically observed LL GRBs and/or shock breakout emission. 
  
For comparison, we also show one of the predictions of the EJ model for $\Gamma=5$. We assume that the luminosity function is constant and the redshift dependence is taken from Ref.~\cite{Porciani:2001kh} but also follows the SFR. Although the model uncertainty is rather large, we confirm the previous results that the EJ model may also give a significant contribution to the diffuse neutrino flux~\cite{Murase:2006mm,Gupta:2006jm} at large obeserved energies (i.e. $E_{\nu,\,obs} \gtrsim 1$ PeV). The spectral shape of earlier results~\cite{Murase:2006mm,Gupta:2006jm} is seen by the recent estimate of Ref. \cite{Tamborra:2015qza}. But the overall normalization is different due to different assumptions on the CR loading factor and LL GRB rate.
In Fig.~\ref{fig:flux}, we show the $p\gamma$ component for the EJ model while the $pp$ component is less important.  

By definition, LL GRBs have gamma-ray counterparts, which are attributed to the shock breakout emission in the CJ-SB model or jet emission in the EJ model. Regardless of the viability of each model as  an explanation for LL GRBs, the existence of choked jets is naturally expected and should be anticipated in any situation with a jet buried deep inside a star with or without extended material around it. In the CJ model, there is no obvious high-energy electromagnetic counterpart. It is known that long GRBs are associated with core collapse SNe (e.g., GRB 060218/SN 2006aj, GRB 980425/SN 1998bw, and GRB 100316D/SN 2010bh). These SNe tend to be Type Ibc meaning little to no hydrogen or helium is observed in the ejecta. There are also broad-line Type Ibc SNe which are often referred to as hypernovae. SNe associated with LL GRBs (although they are not necessarily hypernovae) are also characterized by transreletavistic ejecta. It is then reasonable to assume that a significant fraction of broad-line Type Ibc SNe or hypernovae, even those without accompanying GRBs, contain a choked jet such as in the CJ model. A joint investigation between IC and the ROTSE collaboration attempted to detect optical transients from Type Ibc SNe coincident with neutrino multiplets \cite{Abbasi:2011ja}. No such events were found, but an upper limit on the rate of SNe with a jet was found to be $\lesssim 4.2\%$ of the assumed rate of ccSNe within 10 Mpc. This study could be replicated using neutrino singlets from IC and the All-Sky Automated Survey for Supernovae (ASAS-SN;~\url{www.astronomy.ohio-state.edu/~assassin}) network to further constraint our CJ model. 

\begin{figure}[t]
\includegraphics[width=3.6in]{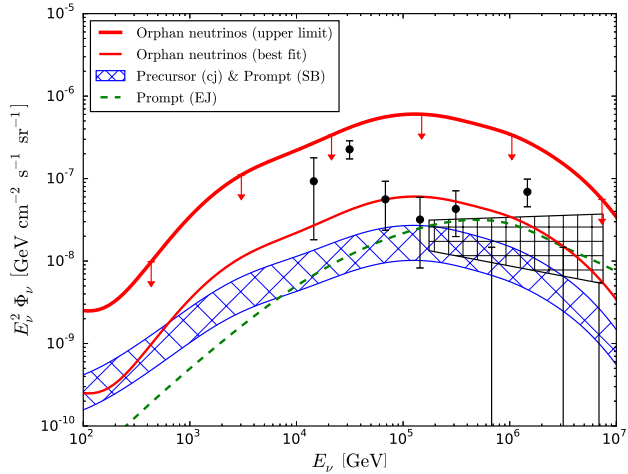}
\caption{All-flavor diffuse neutrino fluxes from choked jets.  Neutrino emission from LL GRBs is shown for the CJ-SB model (this work) and EJ model~\cite{Murase:2006mm}. In addition, orphan neutrino emission from choked jets is included (thick curves).  See the text for details. 
}
\vspace{-1.\baselineskip}
\label{fig:models} 
\end{figure}

More specific results involving the CJ scenario are shown in Fig.~\ref{fig:models}. Although the model uncertainty is large (since $\xi_{\rm CR}$ for GRB jets is not well-known), our results indicate that it is possible for choked jets to achieve the observed level of the diffuse neutrino flux.  In principle, lower values of $\xi_{\rm CR}$ could be compensated by larger values of $f_{\rm cho}$. We set a rough upper limit on the choked jet contribution by using the observed hypernova rate, $R_{\rm HN}\approx4000\,\mathrm{Gpc^{-3}\,yr^{-1}}$~\cite{Liang:2006ci,smith+11ccsne}, which gives $f_{\rm cho}\lesssim40$~\cite{Murase:2014tsa}. A CJ rate similar to that of HNe is also in agreement with Ref.~\cite{Razzaque:2003uv} who considered neutrinos from electromagnetically dim sources. They found that transients with a rate of $\lesssim10^5~{\rm yr}^{-1}$ up to $z\sim1$ could produce a detectable flux of neutrinos.  A similar result is also obtained by Ref.~\cite{2015arXiv151201559T}.

Using the assumed rate and CR energy injection per event, the all-flavor diffuse neutrino flux is analytically estimated to be
\begin{multline}
\label{eq:nuapprox}
E_{\nu}^2\Phi_{\nu}\simeq0.76\times{10}^{-7}~{\rm GeV}~{\rm cm}^{-2}~{\rm s}^{-1}~{\rm sr}^{-1}~ f_{\rm sup}{\rm min}[1,f_{p\gamma}]\\
\times\left(\frac{\xi_z}{3}\right)\left(\frac{f_{\rm cho}{\mathcal E}_{\rm CR}R_{\rm LL}}{{10}^{45}~{\rm erg}~{\rm Mpc}^{-3}~{\rm yr}^{-1}}\right){\mathcal R}_{p,1}^{-1}\,,
\end{multline}
where $f_{\rm sup}$ is the suppression factor due to meson and muon cooling, $\xi_z$ is a factor accounting for redshift evolution of the rate~\cite{Waxman:1998yy,Bahcall:1999yr}, and ${\mathcal R}_p=\ln(\varepsilon_p^{M}/\varepsilon_p^{\rm min})\sim10$ is the bolometric correction factor. 
Interestingly, even this simple-minded calculation is remarkably close to the measured all-flavor diffuse flux of neutrinos~\citep{Aartsen:2015ita}, 
\begin{equation}
E_\nu^2 \Phi_{\nu}^{\rm ob}|_{30~{\rm TeV}}\sim10^{-7}~{\rm GeV}\,{\rm cm}^{-2}\,{\rm s}^{-1}\,{\rm sr}^{-1}.
\end{equation}
Murase et al.~\cite{Murase:2015xka} showed that neutrino sources obscured in the GeV-TeV gamma-ray range are necessary to explain the IceCube data below 100~TeV with extragalactic sources, independently of the neutrino production mechanism. LL GRBs and choked jets satisfy this criterion, and their contribution to the extragalactic gamma-ray background is negligible. 

As a lower limit, for given parameters we can use the rate of observed LL GRBs (i.e., excluding choked jets without prominent shock breakout emission). Noting that the emitted CR energy is roughly the same as that for hypernovae, $R_{\rm LL} \sim 100 \,\mathrm{Gpc^{-3}\,yr^{-1}}$ results in $E_{\nu}^2\Phi_{\nu}\sim{10}^{-8}~{\rm GeV}~{\rm cm}^{-2}~{\rm s}^{-1}~{\rm sr}^{-1}$, which is compatible with the IceCube data above 100~TeV.  Using modest values of $f_{\rm cho}\sim{\rm a~few}$ allows us to reasonably fit the IceCube data obtained from the combined analysis.     

One can set an optimistic upper limit for the contribution of orphan neutrinos from choked jets  by assuming the same spectral shape as in the CJ-SB model with CR energy injection rate of $Q_{\rm CR} \lesssim{10}^{46}\,\mathrm{erg\,Mpc^{-3}\,yr^{-1}}$. This upper limit is set by the reasonable expectation that the CR injection by GRB jets does not exceed the CR injection by SN remnants (see Ref.~\cite{Murase:2015xka}). Note that the CR injection rate inferred by observations of the Galactic CRs is $\sim{10}^{45}-{10}^{46}\,\mathrm{erg\,Mpc^{-3}\,yr^{-1}}$~\cite{Katz:2013ooa}. 
Fig.~\ref{fig:models} indicates the the optimistic upper limit can exceed the IceCube data in principle. The spectral shape is suggestive as it is globally soft for $10\,\mathrm{TeV} \lesssim E_\nu \lesssim 5\,\mathrm{PeV}$, but avoids the constraints set by the {\it Fermi} extragalactic gamma-ray background measurement in the sub-TeV range~\cite{Murase:2013rfa,Senno:2015tra,Ackermann:2014usa}. 

While our results show that the choked jets are energetically plausible as high-energy neutrino sources, we have not tuned parameters to fit the IceCube data quantitatively. Because of the limited statistics of the IceCube data and a tension among the different analyses, such an attempt is beyond the scope of the present  work. Also, better fits of the spectral shape and normalization of the diffuse flux would be possible by changing the parameters of the jet and/or extended material within model uncertainties.

\section{Summary and Discussion} 

We have revised the VHE neutrino emission from LL GRBs, taking into account the jet-stalling condition for a dense circumstellar wind and radiation constraints. Lower-power jets and/or more extended external material are more favorable for both jet-stalling and VHE neutrino production.  This implies the relevance of ``orphan'' neutrinos from choked jets with no prominent electromagnetic counterparts.  Using properties inferred from LL GRB observations (where we did not tune the parameters to explain the IceCube data), we found that the spectral shape and flux normalization of the CJ model can be consistent with the present IceCube data. Although the rate of choked jets with dim shock breakout emission is unknown, it is plausible to use the rate of broad-line Type Ibc SNe as an upper limit. It is currently difficult to exclude parts of the CJ/CJ-SB parameter space in our model since there are many degeneracies (e.g.  the amount of CR energy injected by CJs versus the rate of HNe which have an accompanying luminous, relativistic jet as seen in Eq. \ref{eq:nuapprox}). However, by assuming that the seed photons for $p\gamma$ interactions in the CJ scenario come predominately from the black body emission of the jet head, it should be possible to constrain the combined parameter space of the jet luminosity, head position, and jet Lorentz factor (see \S \ref{sec:nuprod} above) by using the energy at which the observed neutrino flux is at its maximum.  An important prediction of the CJ-SB model is that the majority of neutrinos will be precursors to the prompt gamma-ray emission. Therefore, for a neutrino-LL GRB coincidence search it is imperative to look for neutrinos in temporal blocks $\sim100-1000$~s before the GRB trigger. Based on the currently implied rate of LL GRBs, two such coincident detections can be expected to occur within the next five years of IceCube operation. 
For this purpose, we emphasize that better all-sky monitors in the x-ray and gamma-ray range, which are also suitable for detections of high-redshift GRBs, are necessary. 
Coincident searches can also be expanded to include hypernovae and other energetic SNe. While the rate of such events are higher, the delay between the neutrino and optical/x-ray signal is unknown but may be $\gtrsim1000$ s.  

Detecting precursor neutrinos with short duration would support the CJ-SB model for LL GRBs.
Further observations of $1\,\mathrm{TeV} \lesssim E_\nu \lesssim 100\,\mathrm{TeV}$ neutrinos may provide further information about the envelope mass, radius, and density profile for the extended material around WR stars. Such regions are hard to probe observationally, especially if the mass loss rate of GRB progenitors significantly increases in the months-weeks before collapse.  
While we considered only environments around WR stars in this work, the treatment may be equally important in BSGs, e.g., \cite{meszaros+rees01jets}, which are also believed to be collapsar progenitors and tend to have stellar envelopes that extend up to ${10}^{13.5}\,\mathrm{cm}$. 



\medskip
\acknowledgments
We thank Mauricio Bustamante and Kazumi Kashiyama for useful discussions as well as the anonymous referee for helpful comments.  This work is partially supported by NASA NNX 13AH50G (P. M. and N.S.).  
While this manuscript was being finalized, we became aware of a related but independent work came out on arXiv~\cite{2015arXiv151201559T}. One important difference in our approach is that we take into account the jet-stalling condition and radiation constraints on shock acceleration which are indeed relevant for neutrino production inside dense environments.  
Assuming high-power jets or long durations with $t_{\rm eng}\sim10^6$~s leads to successful jets rather than choked jets, where IceCube constraints should be applied.

\bibliography{kmurase.bib,supbib.bib}

\end{document}